\documentclass[a4paper,english,superscriptaddress,showpacs,showkeys]{revtex4-2}
\usepackage[T2A,T1]{fontenc}
\usepackage[utf8]{inputenc}
\usepackage{color}
\usepackage{babel}
\usepackage{array}
\usepackage{units}
\usepackage{bm}
\usepackage{amsmath}
\usepackage{amssymb}
\usepackage{graphicx}
\usepackage[bookmarks=true,bookmarksnumbered=false,bookmarksopen=true,bookmarksopenlevel=1,
 breaklinks=false,pdfborder={0 0 1},backref=false,colorlinks=true]
 {hyperref}
\hypersetup{
 pdfauthor={S. G. Salnikov},
 pdfborderstyle=,citecolor=blue,urlcolor=blue}

\makeatletter

\pdfpageheight\paperheight
\pdfpagewidth\paperwidth

\DeclareRobustCommand{\cyrtext}{%
  \fontencoding{T2A}\selectfont\def\encodingdefault{T2A}}
\DeclareRobustCommand{\textcyrillic}[1]{\leavevmode{\cyrtext #1}}

\providecommand{\tabularnewline}{\\}

\makeatother

\begin{document}
\title{$N\bar{N}$ production in $e^{+}e^{-}$ annihilation near the threshold
revisited}
\author{A. I. Milstein}
\email{A.I.Milstein@inp.nsk.su}

\affiliation{\textit{Budker Institute of Nuclear Physics, 630090, Novosibirsk,
Russia}}
\affiliation{\textit{Novosibirsk State University, 630090, Novosibirsk, Russia}}
\author{S. G. Salnikov}
\email{S.G.Salnikov@inp.nsk.su}

\affiliation{\textit{Budker Institute of Nuclear Physics, 630090, Novosibirsk,
Russia}}
\affiliation{\textit{Novosibirsk State University, 630090, Novosibirsk, Russia}}
\date{\today}
\begin{abstract}
Production of $p\bar{p}$ and $n\bar{n}$ pairs in $e^{+}e^{-}$ annihilation
near the threshold of the process is discussed with account for the
new experimental data appeared recently. Since a significant part
of these new data was obtained at energies noticeably exceeding the
threshold, we also take into account the form factor describing the
amplitude of $N\bar{N}$ pair production at small distances. The effective
optical potential, which describes a sharp dependence of the $N\bar{N}$
production cross sections near the threshold, consists of the central
potential for $S$ and $D$ waves and the tensor potential. These
potentials differ for the states with isospin $I=0$ and $I=1$ of
$N\bar{N}$ pair. The optical potential describes well $N\bar{N}$
scattering phases, the cross sections of $p\bar{p}$ and $n\bar{n}$
production in $e^{+}e^{-}$ annihilation near the threshold, the electromagnetic
form factors $G_{E}$ and $G_{M}$ for protons and neutrons, as well
as the cross sections of the processes $e^{+}e^{-}\to6\pi$ and $e^{+}e^{-}\to K^{+}K^{-}\pi^{+}\pi^{-}$.
\end{abstract}
\maketitle
\noindent
\global\long\def\im#1{\qopname\relax{no}{Im}#1}%

\section{Introduction}

A strong energy dependence of the cross sections of baryon-antibaryon
and meson-antimeson pair production has been observed in many processes
near the thresholds of the corresponding reactions. Some of these
processes are $e^{+}e^{-}\to p\bar{p}$~\citep{Aubert2006,Lees2013,Akhmetshin2016,Ablikim2020,Ablikim2021b,Akhmetshin2019,Ablikim2015,Ablikim2019},
$e^{+}e^{-}\to n\bar{n}$~\citep{Achasov2014,Ablikim2021f,Achasov2022},
$e^{+}e^{-}\to\Lambda_{(c)}\bar{\Lambda}_{(c)}$~\citep{Aubert2007,Pakhlova2008,Ablikim2018,Ablikim2018b},
$e^{+}e^{-}\to B\bar{B}$~\citep{Aubert2009}, and $e^{+}e^{-}\to\phi\Lambda\bar{\Lambda}$~\citep{Ablikim2021c}.
This anomalous behavior can naturally be explained by small relative
velocities of the produced particles. Therefore, they can interact
strongly with each other for a sufficiently long time. As a result,
the wave function of the produced pair changes significantly (the
so-called final-state interaction). The idea on the final-state interaction
as a source of anomalous energy dependence of the cross sections near
the thresholds has been expressed in many papers~\citep{Dmitriev2006,Haidenbauer2006a,dmitriev2007final,dmitriev2014isoscalar,Haidenbauer2014,Dmitriev2016,Milstein2018,Haidenbauer2016,Milstein2021,Milstein2022a,Milstein2022}.
However, the technical approaches used in these papers were different.
It turned out that in almost all cases the anomalous behavior of the
cross sections is successfully described by the final-state interaction.

Unfortunately, information on the potentials, which are responsible
for the final-state interaction, is very limited. However, instead
of trying to find these potentials from the first principles, one
can use some effective potentials, which are described by a small
number of parameters. These parameters are found by comparison of
the predictions with a large amount of experimental data. Such an
approach has justified itself in all known cases.

One of the most complicated processes for investigation is N$\bar{N}$
pair production in $e^{+}e^{-}$ annihilation near the threshold.
To describe the process, it is necessary to take into account the
central part of the potential for $S$ and $D$ waves and the tensor
part of the potential. In addition, these potentials are different
in the isoscalar and isovector channels. Another circumstance, that
is necessary to take into account, is a large number of $N\bar{N}$
annihilation channels to mesons. As a result, instead of the usual
real potentials, one has to use the so-called optical potentials containing
the imaginary parts. Note that in a narrow region near the thresholds
of $p\bar{p}$ and $n\bar{n}$ production, the Coulomb interaction
of $p$ and $\bar{p}$ should also be taken into account as well as
the proton and neutron mass difference.

The details of approach that allows one to solve the specified problem
are given in our paper~\citep{Milstein2018}. However, in that paper
the parameters of the potentials and the corresponding predictions
for various characteristics of the processes were based on the old
experimental data on the production of $p\bar{p}$ and $n\bar{n}$
pairs. Moreover, a significant part of the uncertainty in the parameters
of the model was related to a poor accuracy of the experimental data
on the cross section of $n\bar{n}$ pair production. Recently, new
data have appeared on $n\bar{n}$ pair production in $e^{+}e^{-}$
annihilation near the threshold~\citep{Ablikim2021f,Achasov2022}.
These data differ significantly from the previous ones and have a
fairly high accuracy compared to the previous experiments. Therefore,
it became necessary to perform a new analysis of the numerous experimental
data within our model.

The approach in Ref.~\citep{Milstein2018} was based on the assumption
that the amplitude of a hadronic system production at small distances
weakly depends on energy of the system near the threshold of the process.
Therefore, in Ref.~\citep{Milstein2018} this amplitude was considered
as energy independent, and strong energy dependence of the cross section
has appeared via the energy dependence of the wave function due to
the final-state interaction. In order to use the new data obtained
at energies significantly above the threshold (but in the non-relativistic
approximation), in the present paper we introduce the phenomenological
dipole form factor which describes the amplitude of a hadronic system
production at small distances.

The aim of the present work is the analysis of $N\bar{N}$ real and
virtual pair production in $e^{+}e^{-}$ annihilation with the new
experimental data taken into account. We show that our model, which
contains a relatively small number of parameters, successfully describes
the energy dependence of $N\bar{N}$ scattering phases (see Ref.~\citep{zhou2012energy}
and references therein), the energy dependence of the cross sections
of $p\bar{p}$ and $n\bar{n}$ pair production near the threshold~\citep{Aubert2006,Lees2013,Akhmetshin2016,Akhmetshin2019,Ablikim2015,Ablikim2019,Ablikim2020,Ablikim2021b,Achasov2014,Ablikim2021f,Achasov2022},
the electromagnetic form factors $G_{E}$ and $G_{M}$ for protons
and neutrons in the time-like region~\citep{Aubert2006,Lees2013,Akhmetshin2016,Ablikim2019,Ablikim2020,Ablikim2021b},
as well as the anomalous behavior of the cross sections of the processes
$e^{+}e^{-}\to6\pi$~\citep{Aubert2006a,Akhmetshin2013,Lukin2015,Akhmetshin2019}
and $e^{+}e^{-}\to K^{+}K^{-}\pi^{+}\pi^{-}$~\citep{Aubert2005,Aubert2007c,Akhmetshin2019}.

\section{Description of the model}

The wave function of the $N\bar{N}$ system produced in $e^{+}e^{-}$
annihilation through one virtual photon contains four components,
namely, $p\bar{p}$ pair in $S$ and $D$ waves and $n\bar{n}$ pair
in $S$ and $D$ waves. It is necessary to take into account $p\bar{p}$
and $n\bar{n}$ pairs together in the wave function due to the charge-exchange
processes $p\bar{p}\leftrightarrow n\bar{n}$. Contributions of $S$
and $D$ waves must be taken into account together due to a tensor
potential, which, for the total angular momentum $J=1$ and the total
spin $s=1$, leads to mixing of states with the orbital angular momenta
$L=0$ and $L=2$. In the absence of the effects violating the isotopic
invariance (the Coulomb $p\bar{p}$ interaction and the proton and
neutron mass difference), the potential in the states with a certain
isospin $I=0,\,1$ has the form
\begin{equation}
V^{I}=V_{S}^{I}(r)\delta_{L0}+V_{D}^{I}(r)\delta_{L2}+V_{T}^{I}(r)\left[6\left(\bm{s}\cdot\bm{n}\right)^{2}-4\right],\label{eq:potential}
\end{equation}
where $\bm{s}$~is the spin operator of $N\bar{N}$ pair ($s=1$),
$\bm{n}=\bm{r}/r$, and $\boldsymbol{r}=\boldsymbol{r}_{N}-\boldsymbol{r}_{\bar{N}}$.
The potentials $V_{S}^{I}(r)$, $V_{D}^{I}(r)$,~and $V_{T}^{I}(r)$
correspond to interaction in the states with $L=0$ and $L=2$, as
well as the tensor interaction. With account for the effects violating
the isotopic invariance we have to solve not two independent systems
for each isospin but one system of equations for the four-component
wave function $\Psi$ (see Ref.~\citep{Milstein2018} for more details)
\begin{align}
 & \begin{aligned} & \left[p_{r}^{2}+\mu\,\mathcal{V}-\mathcal{K}^{2}\right]\Psi=0\,,\qquad &  & \Psi^{T}=\left(u^{p},w^{p},u^{n},w^{n}\right),\\
 & \mathcal{K}^{2}=\begin{pmatrix}k_{p}^{2}\mathbb{I} & 0\\
0 & k_{n}^{2}\mathbb{I}
\end{pmatrix}, &  & \mathbb{I}=\begin{pmatrix}1 & 0\\
0 & 1
\end{pmatrix},
\end{aligned}
\nonumber \\
 & \mu=\frac{1}{2}\left(m_{p}+m_{n}\right),\qquad k_{p}^{2}=\mu E\,,\qquad k_{n}^{2}=\mu(E-2\Delta)\,,\qquad\Delta=m_{n}-m_{p}\,,\label{eq:equation}
\end{align}
where $\Psi^{T}$~denotes a transposition of $\Psi$, $(-p_{r}^{2})$~is
the radial part of the Laplace operator, $u^{p}(r)$, $w^{p}(r)$~and
$u^{n}(r)$, $w^{n}(r)$~are the radial wave functions of $p\bar{p}$
or $n\bar{n}$ pair with $L=0$~and $L=2$, respectively, $m_{p}$~and
$m_{n}$~are the proton and neutron masses, $E$~is the energy of
a system counted from the $p\bar{p}$ threshold, $\hbar=c=1$. In
Eq.~\eqref{eq:equation}, $\mathcal{V}$~is the matrix $4\times4$
which accounts for the $p\bar{p}$ interaction and $n\bar{n}$ interaction
as well as transitions $p\bar{p}\leftrightarrow n\bar{n}$. This matrix
can be written in a block form as
\begin{equation}
\mathcal{V}=\begin{pmatrix}\mathcal{V}^{pp} & \mathcal{V}^{pn}\\
\mathcal{V}^{pn} & \mathcal{V}^{nn}
\end{pmatrix},
\end{equation}
where the matrix elements read
\begin{align}
 & {\cal V}^{pp}=\frac{1}{2}({\cal U}^{1}+{\cal U}^{0})-\frac{\alpha}{r}\,\mathbb{I}+{\cal U}_{cf}\,, &  & {\cal V}^{nn}=\frac{1}{2}({\cal U}^{1}+{\cal U}^{0})+{\cal U}_{cf}\,, &  & {\cal V}^{pn}=\frac{1}{2}({\cal U}^{0}-{\cal U}^{1})\,,\nonumber \\
 & {\cal U}^{I}=\begin{pmatrix}V_{S}^{I} & -2\sqrt{2}\,V_{T}^{I}\\
-2\sqrt{2}\,V_{T}^{I} & \;V_{D}^{I}-2\,V_{T}^{I}
\end{pmatrix}, &  & {\cal U}_{cf}=\frac{6}{\mu r^{2}}\begin{pmatrix}0 & 0\\
0 & 1
\end{pmatrix}\,,
\end{align}
and $\alpha$~is the fine-structure constant and $\mathbb{I}$~is
the unit matrix $2\times2$.

The equation~\eqref{eq:equation} has four linearly independent regular
at $r\to0$ solutions $\Psi_{iR}$ ($i=1\div4$) with asymptotics
at $r\to\infty$ given in~\citep{Milstein2018}. The proton and neutron
electromagnetic form factors are expressed in terms of the components
of these wave functions as follows
\begin{align}
 & G_{M}^{p}=\left\{ g_{p}u_{1R}^{p}(0)+g_{n}u_{1R}^{n}(0)+\frac{1}{\sqrt{2}}\left[\vphantom{\Bigl(\Bigr)}g_{p}u_{2R}^{p}(0)+g_{n}u_{2R}^{n}(0)\right]\right\} F_{D}(q)\,,\nonumber \\
 & G_{E}^{p}=\frac{q}{2\mu}\left\{ g_{p}u_{1R}^{p}(0)+g_{n}u_{1R}^{n}(0)-\vphantom{\frac{1}{\sqrt{2}}}\sqrt{2}\left[\vphantom{\Bigl(\Bigr)}g_{p}u_{2R}^{p}(0)+g_{n}u_{2R}^{n}(0)\right]\right\} F_{D}(q)\,,\nonumber \\
 & G_{M}^{n}=\left\{ g_{p}u_{3R}^{p}(0)+g_{n}u_{3R}^{n}(0)+\frac{1}{\sqrt{2}}\left[\vphantom{\Bigl(\Bigr)}g_{p}u_{4R}^{p}(0)+g_{n}u_{4R}^{n}(0)\right]\right\} F_{D}(q)\,,\nonumber \\
 & G_{E}^{n}=\frac{q}{2\mu}\left\{ g_{p}u_{3R}^{p}(0)+g_{n}u_{3R}^{n}(0)-\vphantom{\frac{1}{\sqrt{2}}}\sqrt{2}\left[\vphantom{\Bigl(\Bigr)}g_{p}u_{4R}^{p}(0)+g_{n}u_{4R}^{n}(0)\right]\right\} F_{D}(q)\,,\nonumber \\
 & F_{D}(q)=\frac{1}{\left(1-\frac{q^{2}}{q_{0}^{2}}\right)^{2}}\,,\qquad q=2\mu+E\,,\qquad q_{0}=\unit[840]{MeV}\,.
\end{align}
Here $F_{D}(q)$ is the phenomenological dipole form factor that takes
into account the energy dependence of the amplitude of the hadronic
system production at small distances, $u_{iR}^{p}(0)$ and $u_{iR}^{n}(0)$
are the energy-dependent components of the wave function at $r=0$,
$g_{p}$ and $g_{n}$ are energy-independent fitting parameters.

The cross sections of $p\bar{p}$ and $n\bar{n}$ pair production,
which we refer to as the elastic cross sections, have the form
\begin{align}
 & \sigma_{\mathrm{el}}^{p}=\frac{4\pi k_{p}\alpha^{2}}{q^{3}}F_{D}^{2}(q)\left[\left|g_{p}u_{1R}^{p}(0)+g_{n}u_{1R}^{n}(0)\right|^{2}+\left|g_{p}u_{2R}^{p}(0)+g_{n}u_{2R}^{n}(0)\right|^{2}\right],\nonumber \\
 & \sigma_{\mathrm{el}}^{n}=\frac{4\pi k_{n}\alpha^{2}}{q^{3}}F_{D}^{2}(q)\left[\left|g_{p}u_{3R}^{p}(0)+g_{n}u_{3R}^{n}(0)\right|^{2}+\left|g_{p}u_{4R}^{p}(0)+g_{n}u_{4R}^{n}(0)\right|^{2}\right].
\end{align}
In the absence of the final-state interaction, we have $u_{1R}^{p}(0)=u_{3R}^{n}(0)=1$,
and the rest $u_{iR}^{p}(0)$ and $u_{iR}^{n}(0)$ vanish. The functions
$u_{3R}^{p}(0)$ and $u_{1R}^{n}(0)$ differ from zero due to the
charge-exchange process, while nonzero values of $u_{2R}^{p}(0)$,
$u_{2R}^{n}(0)$, $u_{4R}^{p}(0)$, and $u_{4R}^{n}(0)$ are the consequence
of the tensor forces. Note that $\left|G_{E}^{p}/G_{M}^{p}\right|$
and $\left|G_{E}^{n}/G_{M}^{n}\right|$ differ from unity solely due
to the tensor forces. For $E=0$ these ratios are equal to unity,
since at the threshold the contribution of the $D$ wave vanishes.

\setlength{\tabcolsep}{1em}

\begin{table}
\begin{centering}
\begin{tabular}{|l|c|c|c|c|c|c|}
\hline
 & $\widetilde{U}_{S}^{0}$ & $\widetilde{U}_{D}^{0}$ & $\widetilde{U}_{T}^{0}$ & $\widetilde{U}_{S}^{1}$ & $\widetilde{U}_{D}^{1}$ & $\widetilde{U}_{T}^{1}$\tabularnewline
\hline
$U_{i}\,(\mathrm{MeV})$ & $-196$ & $80.8$ & $-2.2$ & $-36.3$ & $401.6$ & $15.2$\tabularnewline
$W_{i}\,(\mathrm{MeV})$ & $167.3$ & $225.4$ & $-2$ & $-16.4$ & $217.2$ & $1.5$\tabularnewline
$a_{i}\,(\mathrm{fm})$ & $0.701$ & $1.185$ & $2.704$ & $1.294$ & $0.739$ & $1.289$\tabularnewline
\hline
$g_{i}$ & \multicolumn{3}{c|}{$g_{p}=14.1$} & \multicolumn{3}{c|}{$g_{n}=3.6-1.1i$}\tabularnewline
\hline
\end{tabular}
\par\end{centering}
\caption{The parameters of the model.}\label{tab:fit}
\end{table}

In addition to the strong energy dependence of the cross sections
$\sigma_{\mathrm{el}}^{p}$ and $\sigma_{\mathrm{el}}^{n}$ near the
threshold, a strong energy dependence reveals also in the cross sections
of meson production in $e^{+}e^{-}$ annihilation near the $N\bar{N}$
pair production threshold~\citep{Aubert2006a,Akhmetshin2013,Lukin2015,Akhmetshin2019,Aubert2005,Aubert2007c}.
Such a behavior is related to the production of virtual $N\bar{N}$
pair below and above the threshold with the subsequent annihilation
of this pair into mesons. Since the probability of virtual $N\bar{N}$
pair production strongly depends on energy, then the probability of
meson production through the intermediate $N\bar{N}$ state also strongly
depends on energy. Meanwhile, the probability of meson production
through other mechanisms has weak energy dependence near the $N\bar{N}$
threshold. To find the cross section $\sigma_{\mathrm{in}}^{I}$ of
meson production through $N\bar{N}$ intermediate state (the inelastic
cross section) with a certain isospin~$I$, one can use the optical
theorem. Due to this theorem, the cross sections $\sigma_{\mathrm{tot}}^{I}=\sigma_{\mathrm{el}}^{I}+\sigma_{\mathrm{in}}^{I}$
are expressed via the imaginary part of the Green's function ${\cal D}\left(r,r'|E\right)$
of the Schrödinger equation:
\begin{align}
 & \sigma_{\mathrm{tot}}^{I}=\frac{2\pi\alpha^{2}}{q^{3}}F_{D}^{2}(q)\im{\left[\left(\mathcal{G}^{I}\right)^{\dagger}{\cal D}\left(0,0|E\right)\mathcal{G}^{I}\right]},\nonumber \\
 & \left(\mathcal{G}^{0}\right)^{T}=\frac{g_{p}+g_{n}}{2}\cdot\left(1,\,0,\,1,\,0\right),\qquad\left(\mathcal{G}^{1}\right)^{T}=\frac{g_{p}-g_{n}}{2}\cdot\left(1,\,0,\,-1,\,0\right).
\end{align}
The cross sections $\sigma_{\mathrm{el}}^{I}$ have the form
\begin{multline}
\sigma_{\mathrm{el}}^{0}=\frac{4\pi k_{p}\alpha^{2}}{q^{3}}F_{D}^{2}(q)\left|\frac{g_{p}+g_{n}}{2}\right|^{2}\left[\left|u_{1R}^{p}(0)+u_{1R}^{n}(0)\right|^{2}+\left|u_{2R}^{p}(0)+u_{2R}^{n}(0)\right|^{2}\right]\\
\shoveright{+\frac{4\pi k_{n}\alpha^{2}}{q^{3}}F_{D}^{2}(q)\left|\frac{g_{p}+g_{n}}{2}\right|^{2}\left[\left|u_{3R}^{p}(0)+u_{3R}^{n}(0)\right|^{2}+\left|u_{4R}^{p}(0)+u_{4R}^{n}(0)\right|^{2}\right],}\\
\shoveleft{\sigma_{\mathrm{el}}^{1}=\frac{4\pi k_{p}\alpha^{2}}{q^{3}}F_{D}^{2}(q)\left|\frac{g_{p}-g_{n}}{2}\right|^{2}\left[\left|u_{1R}^{p}(0)-u_{1R}^{n}(0)\right|^{2}+\left|u_{2R}^{p}(0)-u_{2R}^{n}(0)\right|^{2}\right]}\\
+\frac{4\pi k_{n}\alpha^{2}}{q^{3}}F_{D}^{2}(q)\left|\frac{g_{p}-g_{n}}{2}\right|^{2}\left[\left|u_{3R}^{p}(0)-u_{3R}^{n}(0)\right|^{2}+\left|u_{4R}^{p}(0)-u_{4R}^{n}(0)\right|^{2}\right].
\end{multline}
The Green's function satisfies the equation,
\begin{equation}
\left[p_{r}^{2}+\mu\mathcal{V}-\mathcal{K}^{2}\right]{\cal D}\left(r,r'|E\right)=\frac{1}{rr'}\delta\left(r-r'\right),
\end{equation}
and is expressed in terms of regular and irregular solutions of the
Schrödinger equation~\eqref{eq:equation} (see Ref.~\citep{Milstein2018}
for details).

\begin{figure}[t]
\begin{centering}
\includegraphics[width=1\textwidth]{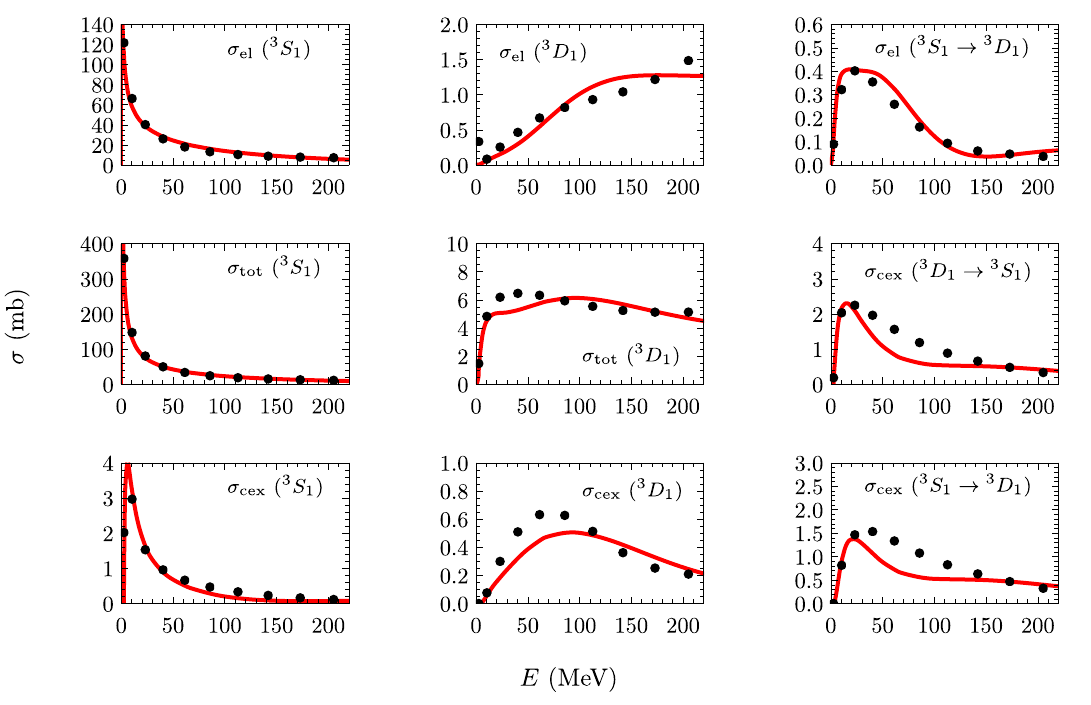}
\par\end{centering}
\caption{The predictions for the cross sections of $p\bar{p}$ scattering
compared with the Nijmegen data~\citep{zhou2012energy}.}\label{fig:Nijmegen}
\end{figure}

\section{Results and Discussion}

The optical potentials $V(r)$ in Eq.~\eqref{eq:potential} are expressed
in terms of the potentials $\widetilde{U}^{0}(r)$ and $\widetilde{U}^{1}(r)$
associated with isoscalar and isovector exchange,
\begin{equation}
V(r)=\widetilde{U}^{0}(r)+\left(\bm{\tau}_{1}\cdot\bm{\tau}_{2}\right)\widetilde{U}^{1}(r)\,,
\end{equation}
where $\bm{\tau}_{1,2}$~are isospin Pauli matrices for nucleon and
antinucleon, respectively. Therefore, $V_{S,D,T}^{I}$ in Eq.~\eqref{eq:potential}
have the form
\begin{align}
 & V_{i}^{1}(r)=\widetilde{U}_{i}^{0}(r)+\widetilde{U}_{i}^{1}(r)\,, &  & V_{i}^{0}(r)=\widetilde{U}_{i}^{0}(r)-3\widetilde{U}_{i}^{1}(r)\,, &  & i=S,D,T\,.
\end{align}
In our model, we use the simplest parametrization of the potentials
$\widetilde{U}^{I}(r)$,
\begin{align}
 & \widetilde{U}_{i}^{0}(r)=\left(U_{i}^{0}-i\,W_{i}^{0}\right)\,\theta(a_{i}^{0}-r)\,,\nonumber \\
 & \widetilde{U}_{i}^{1}(r)=\left(U_{i}^{1}-i\,W_{i}^{1}\right)\,\theta(a_{i}^{1}-r)+U_{i}^{\pi}(r)\,\theta(r-a_{i}^{1})\,, &  & i=S,D,T\,,
\end{align}
where $\theta(x)$~is the Heaviside function, $U_{i}^{I}$, $W_{i}^{I}$,
$a_{i}^{I}$~are free real parameters fixed by fitting the experimental
data, and $U_{i}^{\pi}(r)$~are the terms in the pion-exchange potential
(see, e.g.,~\citep{Ericson1988}).

To fit the parameters of our model, we use the following experimental
data: $N\bar{N}$ scattering phases obtained by the Nijmegen group
(see Ref.~\citep{zhou2012energy} and references therein), the cross
sections of $p\bar{p}$ and $n\bar{n}$ production near the threshold~\citep{Lees2013,Akhmetshin2016,Akhmetshin2019,Ablikim2020,Ablikim2021b,Ablikim2021f,Achasov2022},
modules of electromagnetic form factors $\left|G_{E}^{p}\right|$
and $\left|G_{M}^{p}\right|$~\citep{Ablikim2020}, as well as the
ratios $\left|G_{E}^{p}/G_{M}^{p}\right|$~\citep{Lees2013,Akhmetshin2016,Ablikim2019,Ablikim2020,Ablikim2021b}
and $\left|G_{E}^{n}/G_{M}^{n}\right|$~\citep{Achasov2022}. The
resulting values of parameters are given in Table~\ref{tab:fit}.
For these parameters we obtain $\chi^{2}/N_{df}=98/85$, where $N_{df}$
is the number of degrees of freedom.

\begin{figure}[tb]
\includegraphics[totalheight=5.2cm]{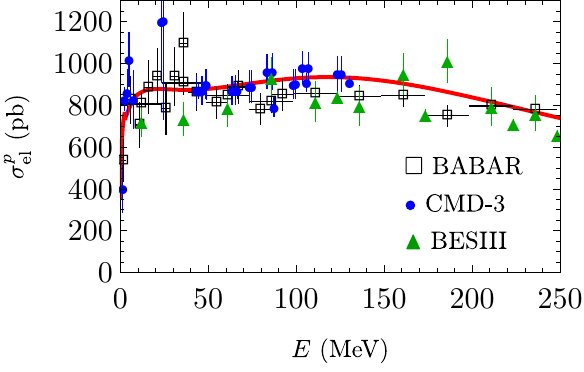}\hfill{}\includegraphics[totalheight=5.2cm]{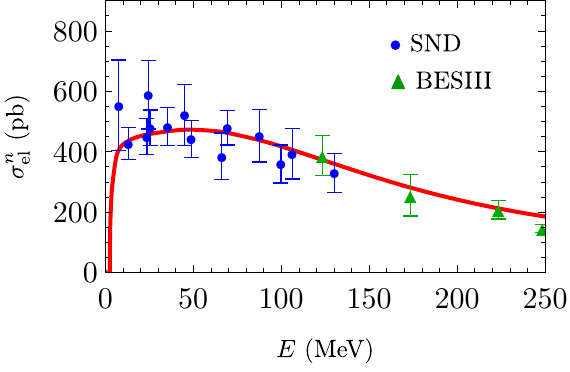}

\includegraphics[totalheight=5.2cm]{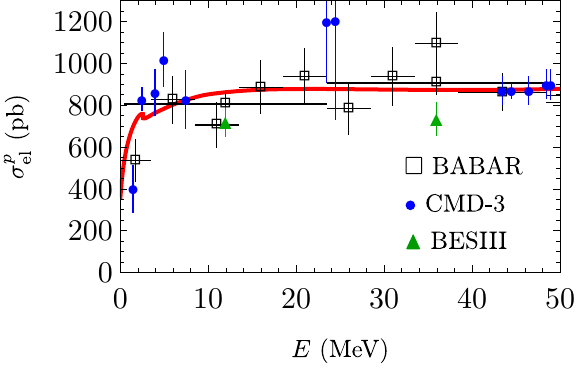}\hfill{}\includegraphics[totalheight=5.2cm]{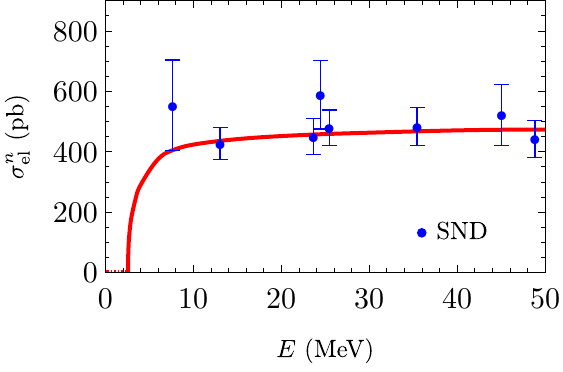}

\caption{The energy dependence of the cross sections of $p\bar{p}$ (left)
and $n\bar{n}$ (right) pair production in $e^{+}e^{-}$ annihilation.
The near-threshold energy region is shown in more details in the bottom
row. The experimental data are taken from BABAR~\citep{Lees2013},
CMD-3~\citep{Akhmetshin2016,Akhmetshin2019}, SND~\citep{Achasov2022},
and BESIII~\citep{Ablikim2020,Ablikim2021b,Ablikim2021f}.}\label{fig:Prod}
\end{figure}

\begin{figure}[!tbph]
\includegraphics[totalheight=5.6cm]{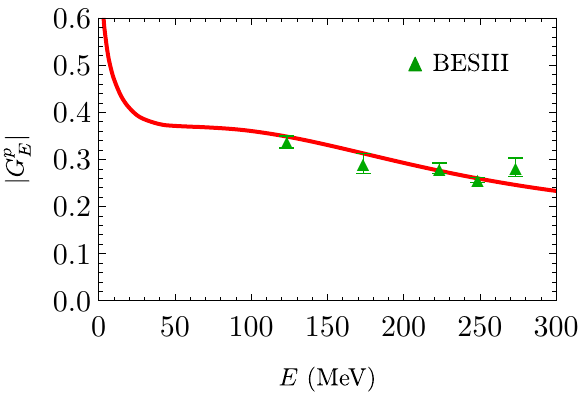}\hfill{}\includegraphics[totalheight=5.6cm]{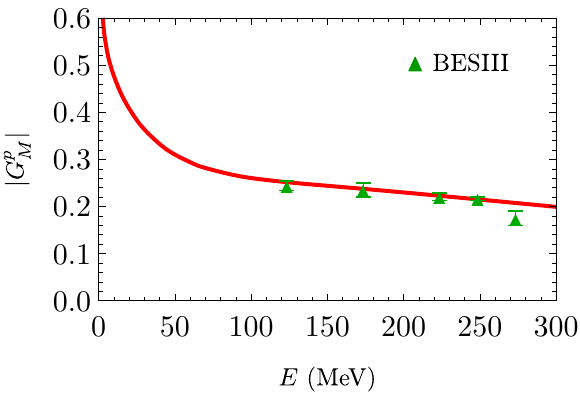}

\includegraphics[totalheight=5.6cm]{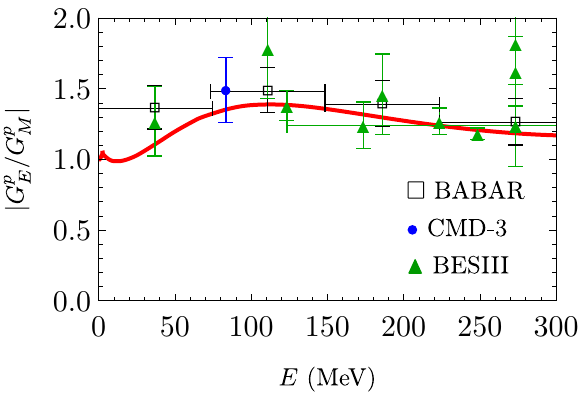}\hfill{}\includegraphics[totalheight=5.6cm]{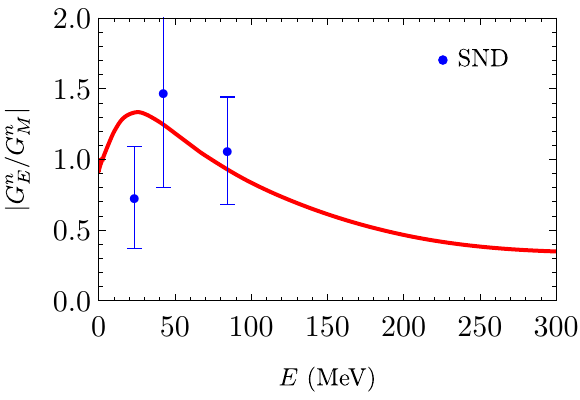}

\caption{The energy dependence of the form factors $\left|G_{E}^{p}\right|$
and $\left|G_{M}^{p}\right|$, as well as the ratios $\left|G_{E}^{p}/G_{M}^{p}\right|$
and $\left|G_{E}^{n}/G_{M}^{n}\right|$. The experimental data are
taken from BABAR~\citep{Lees2013}, CMD-3~\citep{Akhmetshin2016},
SND~\citep{Achasov2022}, and BESIII~\citep{Ablikim2020,Ablikim2021b,Ablikim2019}.}\label{fig:GeGm}
\end{figure}

Fig.~\ref{fig:Nijmegen} shows a comparison of our predictions for
partial cross sections of $p\bar{p}$ scattering with the results
of partial wave analysis~\citep{zhou2012energy}. Fig.~\ref{fig:Prod}
shows the energy dependence of $p\bar{p}$ and $n\bar{n}$ pair production
cross sections. Fig.~\ref{fig:GeGm} shows $\left|G_{E}^{p}\right|$
and $\left|G_{M}^{p}\right|$, as well as the ratios $\left|G_{E}^{p}/G_{M}^{p}\right|$
and $\left|G_{E}^{n}/G_{M}^{n}\right|$. Good agreement of the predictions
with the available experimental data is seen everywhere.

\begin{figure}[!tbph]
\includegraphics[totalheight=5.6cm]{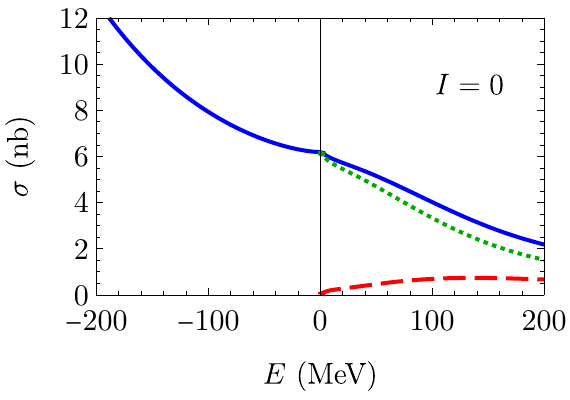}\hfill{}\includegraphics[totalheight=5.6cm]{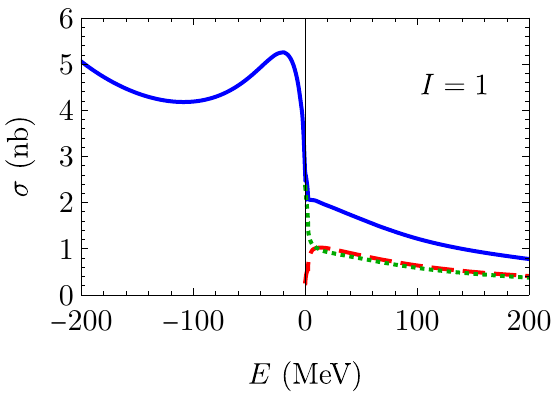}

\caption{The energy dependence of the cross sections $\sigma_{\mathrm{tot}}^{I}$
(solid line), $\sigma_{\mathrm{el}}^{I}$ (dashed line), and $\sigma_{\mathrm{in}}^{I}$
(dotted line) for isospins $I=0,1$.}\label{fig:Tot}
\end{figure}

As mentioned above, the optical theorem allows one to predict the
contributions $\sigma_{\mathrm{in}}^{I}$ to the cross sections of
meson production in $e^{+}e^{-}$ annihilation associated with the
$N\bar{N}$ pairs in an intermediate state. In Fig.~\ref{fig:Tot}
the cross sections $\sigma_{\mathrm{tot}}^{I}$, $\sigma_{\mathrm{el}}^{I}$,
and $\sigma_{\mathrm{in}}^{I}$ are shown. It can be seen that in
the channel with $I=1$ there is a large dip in the cross section
$\sigma_{\mathrm{in}}^{1}$ at the threshold of real $N\bar{N}$ pair
production. At the same time, in the channel with $I=0$ this dip
is practically invisible.

A dip was found in the cross sections of the processes $e^{+}e^{-}\to3\left(\pi^{+}\pi^{-}\right)$~\citep{Aubert2006a,Akhmetshin2013,Akhmetshin2019},
$e^{+}e^{-}\to2\left(\pi^{+}\pi^{-}\pi^{0}\right)$~\citep{Aubert2006a,Lukin2015},
and $e^{+}e^{-}\to K^{+}K^{-}\pi^{+}\pi^{-}$~\citep{Aubert2005,Aubert2007c,Akhmetshin2019}.
Since in our approach we cannot predict the cross sections in each
channel, for comparison of our predictions with experimental data
we use the following procedure. We assume that strong energy dependence
of the cross sections for the production of mesons in each channel
near the $N\bar{N}$ threshold is related to a strong energy dependence
of the amplitude of virtual $N\bar{N}$ pair production in an intermediate
state. We also suppose that the amplitudes of virtual $N\bar{N}$
pair transitions to specific meson states weakly depend on energy
near the threshold of $N\bar{N}$ production. Evidently, other contributions
to meson production cross sections, which are not related to $N\bar{N}$
in an intermediate state, have also a weak energy dependence. Therefore,
we approximate the cross section $\sigma_{\mathrm{mesons}}^{I}$ of
meson production in a state with a certain isospin by the function
\begin{equation}
\sigma_{\mathrm{mesons}}^{I}=a\cdot\sigma_{\mathrm{in}}^{I}+b\cdot E^{2}+c\cdot E+d\,,\label{eq:mesons}
\end{equation}
where $a$, $b$, $c$ \textcyrillic{и} $d$~are some fitting parameters,
which depend on the specific final states.

\begin{figure}[!tbph]
\includegraphics[totalheight=5.6cm]{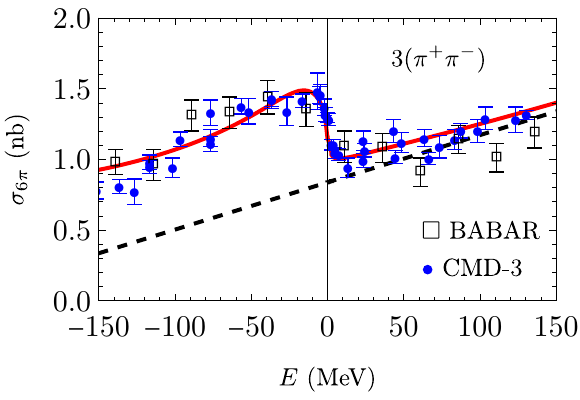}\hfill{}\includegraphics[totalheight=5.6cm]{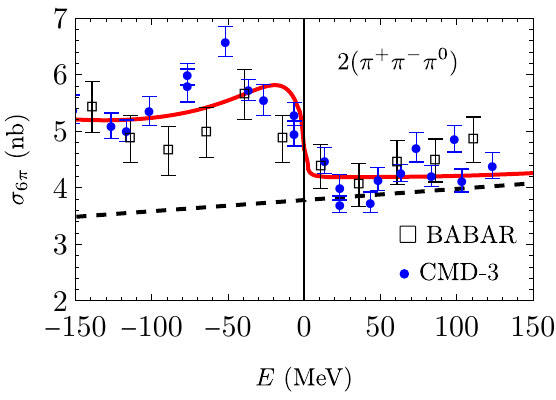}
\begin{centering}
\includegraphics[totalheight=5.6cm]{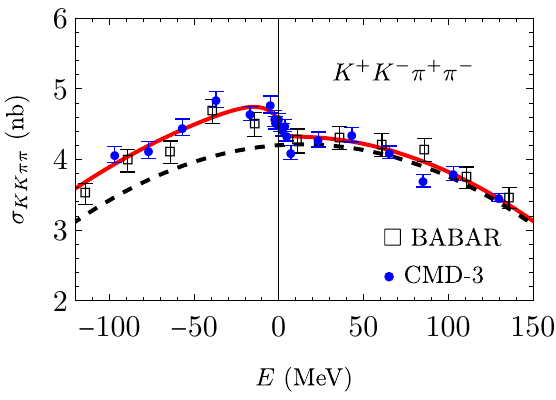}
\par\end{centering}
\caption{The energy dependence of the cross sections for the processes $e^{+}e^{-}\to3\left(\pi^{+}\pi^{-}\right)$,
$e^{+}e^{-}\to2\left(\pi^{+}\pi^{-}\pi^{0}\right)$, and $e^{+}e^{-}\to K^{+}K^{-}\pi^{+}\pi^{-}$.
The experimental data are taken from Refs.~\citep{Aubert2006a,Akhmetshin2013,Akhmetshin2019},
\citep{Aubert2006a,Lukin2015}, and~\citep{Aubert2005,Aubert2007c,Akhmetshin2019},
respectively.}\label{fig:mesons}
\end{figure}

The $6\pi$ final state has isospin $I=1$ due to $G$-parity conservation.
Comparison of our predictions for the $6\pi$ production cross section
with the experimental data is shown in Fig.~\ref{fig:mesons}. For
these processes the fit shows that we can set $b=0$, and the remaining
parameters are $a=0.14$, $c=\unit[3.3\cdot10^{-3}]{nb/MeV}$, $d=\unit[0.84]{nb}$
for $3\left(\pi^{+}\pi^{-}\right)$ production and $a=0.4$, $c=\unit[2\cdot10^{-3}]{nb/MeV}$,
$d=\unit[3.8]{nb}$ for $2\left(\pi^{+}\pi^{-}\pi^{0}\right)$ case.
It can be seen that there is good agreement between our predictions
and experimental data.

Consider now the process $e^{+}e^{-}\to K^{+}K^{-}\pi^{+}\pi^{-}$.
Unlike the $6\pi$ state, the state $K^{+}K^{-}\pi^{+}\pi^{-}$ may
be in both isospin states, $I=1$ and $I=0$. Since our calculations
show that the cross section $\sigma_{\mathrm{in}}^{0}$ has no sharp
energy dependence near the $N\bar{N}$ threshold, then the contribution
of state with $I=0$ can be taken into account in the parameters $b$,
$c$, and $d$. Thus, we can compare the cross section of the process
$e^{+}e^{-}\to K^{+}K^{-}\pi^{+}\pi^{-}$ with formula~\eqref{eq:mesons}
for $I=1$. The fitting parameters for this process are $a=0.11$,
$b=\unit[-6.1\cdot10^{-5}]{nb/MeV^{2}}$, $c=\unit[1.7\cdot10^{-3}]{nb/MeV}$,
$d=\unit[4.2]{nb}$. Comparison of our predictions with experimental
data is also shown in Fig.~\ref{fig:mesons}. Again, there is good
agreement of our predictions and experimental results.

\section{Conclusion}

Using new experimental data on the production of $p\bar{p}$ and $n\bar{n}$
pairs in $e^{+}e^{-}$ annihilation, a simple model is suggested that
successfully describes the cross sections of a few processes with
production of real or virtual $N\bar{N}$ pairs. These processes are
$e^{+}e^{-}\to p\bar{p}$, $e^{+}e^{-}\to n\bar{n}$, $e^{+}e^{-}\to6\pi$,
and $e^{+}e^{-}\to K^{+}K^{-}\pi^{+}\pi^{-}$ near the $N\bar{N}$
production threshold. Moreover, this model describes well the energy
dependence of partial cross sections for nucleon-antinucleon scattering
in states with $L=0,\,2$, $s=1$ and $J=1$, as well as the electromagnetic
form factors of proton and neutron in the time-like region. Since
new experimental data were obtained at energies noticeably exceeding
the $N\bar{N}$ production threshold, an effective dipole form factor
was introduced. It accounts for the energy dependence of the amplitude
of real or virtual $N\bar{N}$ pair production at small distances.
Since the new data on $n\bar{n}$ production have noticeably better
accuracy compared to the previous ones, our predictions became more
accurate. The analysis of meson production in different channels shows
that the strong energy dependence of the meson production cross sections
near the $N\bar{N}$ threshold is related solely to a strong energy
dependence of the amplitude of virtual $N\bar{N}$ pair production
in an intermediate state.

\end{document}